\documentclass[sigconf,screen]{acmart}

\usepackage[final, commandnameprefix=ifneeded]{changes}

\AtBeginDocument{%
  }

\setcopyright{acmlicensed}
\copyrightyear{2026}
\acmYear{2026}
\acmDOI{XXXXXXX.XXXXXXX}
\acmConference[Conference acronym 'XX]{}{**,
  20**}{Woodstock, NY}
\acmISBN{978-1-4503-XXXX-X/2026/06}
\usepackage{tikz}
\usepackage{amsmath}
\usepackage{tabularx}
\usepackage{booktabs}
\usepackage{hyperref}
\usepackage[table]{xcolor}
\usepackage{subcaption}
\usepackage{forest}
\usepackage{graphicx}
\usetikzlibrary{trees, positioning,shapes,shadows,arrows.meta,calc}
\usepackage{makecell}
\usepackage{pgf-pie}
\usepackage{xparse}
\usepackage{etoolbox}
\usepackage{balance}
\usepackage{booktabs}

\usepackage{pgfplots}
\usepackage{pgf-pie}
\pgfplotsset{compat=1.18}
\usepackage{xkeyval}
\usepackage{ifthen}
\usepackage{listings}
\usetikzlibrary{calc}
\usepackage{colortbl}
\usepackage[most]{tcolorbox}
\usepackage{amsthm}
\usepackage{xcolor}
\usepackage{mdframed}
\usepackage{pgfplots}
\pgfplotsset{compat=1.18}
\usepackage{multirow}
\usepackage{threeparttable}
\usepackage{float}

\usepackage{soul}

\def\fig {Figure~}

\def\sec {Section~}

\usepackage{paralist}
\lstdefinestyle{inlinecode}{basicstyle={\ttfamily\scriptsize\bfseries}}

\newcommand{\urls}[1]{{\scriptsize\url{#1}}}

\newcommand{\rlnk}[1]{\href{https://www.reddit.com/comments/#1/}{#1}}
\newcommand{\afif}[1]{\noindent\textcolor{red}{[Afif: {#1}]}}
\newcommand{\gias}[1]{\noindent\textcolor{red}{[Gias: {#1}]}}

\newcommand{\song}[1]{\noindent\textcolor{purple}{~[Song : {#1}]}}
\newcounter{observation}

\newcommand{\observation}[1]{%
  \refstepcounter{observation}%
  \begin{mdframed}[
    linewidth=0.3pt,
    roundcorner=1pt,
    innertopmargin=1pt,
    innerbottommargin=1pt,
    innerleftmargin=1pt,
    innerrightmargin=2pt,
    skipabove=2pt,
    skipbelow=2pt
  ]
  \noindent
  \colorbox{black!80}{%
    \textcolor{white}{\textbf{O\#~\theobservation}}%
  }\hspace{0.5em}%
  #1
  \end{mdframed}
}

\newcommand{\observationl}[1]{%
  \refstepcounter{observation}%
  \begin{mdframed}[
    linewidth=0.4pt,
    roundcorner=2pt,
    innertopmargin=4pt,
    innerbottommargin=4pt,
    innerleftmargin=6pt,
    innerrightmargin=6pt,
    skipabove=6pt,
    skipbelow=6pt
  ]
  \noindent
  \colorbox{black}{%
    \textcolor{white}{\textbf{Observation\#~\theobservation}}%
  }\hspace{0.5em}%
  #1
  \end{mdframed}
}

\newcounter{rec}

\usepackage[table]{xcolor}

\newtcolorbox{takeaway}[1][]{
    colback=blue!10!white,    
    colframe=black,             
    rounded corners,            
    arc=5pt,                   
    boxrule=1pt,               
    left=5pt,                 
    right=5pt,                
    top=1pt,                   
    bottom=1pt, 
    before skip=2pt, 
    after skip=2pt,  
    #1                         
}

\newcounter{rtext}

\usepackage{balance}
\newcommand{\commentout}[1]{}

\makeatletter
\newcommand{\reddittext}[2]{%
  ``\textit{\textcolor{black!85}{#1}}''--#2 
  \@afterindentfalse\@afterheading
}

\makeatletter
\renewenvironment{quote}
  {\list{}{\leftmargin=0.5em \rightmargin=0.25em}%
   \item\relax}
  {\endlist}
\makeatother

\begin{document}

\title{\textit{``Impossible to hide secret ...''}:\\ Uncovering Security and Privacy Issues in LLM-native IDEs}

\author{Mostafijur Rahman Akhond}
\affiliation{%
  \institution{York University}
  \city{Toronto}
  \state{ON}
  \country{CANADA}
}
\email{mostafij@yorku.ca}
\author{Md Afif Al Mamun}
\affiliation{%
  \institution{University of Calgary}
  \city{Calgary}
  \state{AB}
  \country{CANADA}
}
\email{afif.mamun@ucalgary.ca}

\author{Gias Uddin}
\affiliation{%
  \institution{York University}
  \city{Toronto}
  \state{ON}
  \country{CANADA}
}
\email{guddin@yorku.ca}

\author{Song Wang}
\affiliation{%
  \institution{York University}
  \city{Toronto}
  \state{ON}
  \country{CANADA}
}
\email{wangsong@yorku.ca}

\renewcommand{\shortauthors}{Akhond et al.}

\begin{abstract}
LLM-native IDEs (Integrated Development Environments), aka LIDEs, are designed from the ground up to work with Large Language Models (LLMs). LIDEs have found remarkable success in Software Engineering (SE) tasks such as coding, debugging, and program comprehension. 
LIDEs are software systems, and, like any system, they can exhibit vulnerabilities. In this paper, we study the security and privacy issues that developers reported while using popular LIDEs in their development tasks. We collected \added{1.1M} posts from 29 popular subreddits related to LIDEs. We identified \added{446} posts and analyzed over \added{6K} comments to the posts that discussed security and privacy issues in almost all popular LIDEs, such as Cursor, Copilot, Codex, etc. 
Using a mix of qualitative and quantitative methods, we constructed a taxonomy of the reported security and privacy issues. Our results show that most issues in LIDEs stem from system-level design choices, rather than the underlying LLMs, such as user data access, unchecked autonomous actions, etc. To overcome these issues, developers frequently relied on external safeguards like code sandboxing and manual reviewing, highlighting prevalent mistrust among developers about LIDEs. We share lessons from our study to support future design of secure and privacy-aware LIDEs.
\end{abstract}





\maketitle
\balance

\begin{figure*}
    \centering
    \includegraphics[width=0.7\linewidth]{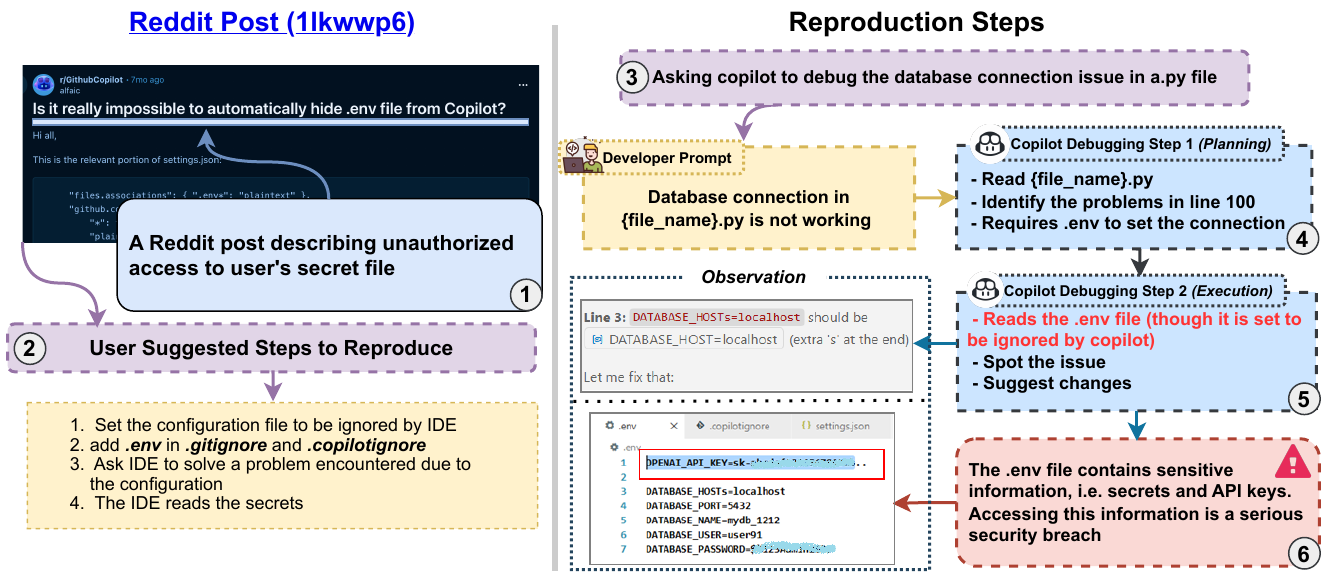}
    \caption{A Reddit post reported unauthorized access to a secret (.env) file (left). We confirm that it still persists (right).}
    \label{fig:motivation}
\end{figure*}

\section{Introduction}

Large Language Models (LLMs) have rapidly transformed the landscape of software development~\cite{hou2024large}. Beyond standalone chat-based assistants, LLMs are now deeply embedded into development workflows through LLM-native Integrated Development Environments (IDEs), aka LIDEs. Unlike traditional IDEs that merely integrate external tools or plugins~\cite{radenkovic2025application}, LIDEs are designed from the ground up to work in tight collaboration with LLMs, enabling capabilities such as real-time code generation, debugging assistance, automated refactoring, and program comprehension. Popular LIDEs, including Cursor, GitHub Copilot, Claude Code, and Codex, have been widely adopted and have demonstrated remarkable success in supporting a broad range of software engineering (SE) tasks. 

However, because LIDEs are built on machine learning systems, they may inherit security and privacy vulnerabilities that can arise both from the underlying LLMs or the system itself \cite{MLSysTechnicalDebt}. 


\commentout{\afif{I was wondering how much of LIDE system is ML? When most of the tasks are tackled in a remote server the LIDE itself just works as medium}}


Prior research on LLM-assisted coding has largely focused on correctness, performance, and isolated vulnerabilities in LLM-generated code snippets~\cite{tong2024codejudge,mathews2024codegen-tests,basic2024large}. Several empirical studies have shown that LLMs frequently generate insecure code patterns, violate secure coding practices, or hallucinate APIs and dependencies when prompts are underspecified~\cite {pearce2022asleep}. Other work has demonstrated that LLM-generated code may reproduce vulnerable patterns present in training data or omit essential validation and access-control logic~\cite{Fu2023SecurityWO}. 
However, these studies typically evaluate LLMs as standalone generators and therefore fail to capture the additional risks that emerge in LIDEs, where LLMs are embedded within complex systems involving tool invocation, persistent state, user data access, and autonomous actions.

In this paper, we focused on understanding the security and privacy issues in LIDEs by empirically studying online developer discussions in \href{https://www.reddit.com/}{Reddit}. 
Reddit offers an unfiltered version of developers' posts and comments, and a unique view of their actual usage of LIDEs. For example, Figure~\ref{fig:motivation} illustrates an incident reported by a Reddit user. The user configured a development environment with GitHub Copilot and managed secrets using an \emph{.env} file. Although access restrictions were configured to prevent GitHub Copilot from accessing the \emph{.env} file, these rules were not properly enforced. We attempted to replicate this issue, and during the process, we observed that after two debugging cycles (i.e., consecutive follow-up prompts) of a database connection issue, GitHub Copilot read and attempted to modify the \emph{.env} file, ignoring the configured access restrictions. This behavior raises serious security and privacy issues, as the file contained private secrets and API keys, exposing developers to significant confidentiality risks. 

These findings motivated us to study (1) the security and privacy issues in popular LIDEs reported by developers in social forums like Reddit, (2) how they are currently mitigating the issues with or without help from the LIDEs designers, and (3) what suggestions we can offer to the LIDEs designers to make their applications more security- and privacy-aware.

For our study, we first collected \added{1.1M} posts from 29 manually curated subreddits that focused on LLM-assisted programming. Each subreddit explicitly used one or more popular LIDEs (e.g., r/cursor for the Cursor IDE). Second, we filtered posts using an LLM-based binary classifier targeting security and privacy discussion, followed by manual human verification, resulting in \added{446} posts and \added{6K} associated comments.
Each filtered post reported a real security/privacy issue a developer experienced while using an LIDE. The comments contained replies and suggestions from other developers on how to address the issue. 
Third, we labeled each post to identify the nature of the reported issues (e.g., unauthorized access to secrets, operational safety). This step produced a taxonomy of security issues and privacy issues in LIDEs. Fourth, we studied the suggestions discussed in the comments. \commentout{This step produced \song{how to produce?} a list of 13 different `mitigation styles' that developers discussed to fix the observed issues.} These mitigation styles were identified by using an AI-assisted approach to extract actionable suggestions from comments, followed by manual verification and consolidation of semantically similar suggestions.\commentout{ Fifth, we produce \song{how to produce?} six recommendations that can be used by the IDE designers and users to design and use more secure LIDEs.} Fifth, based on the mitigation styles discussed by developers and the issues observed in the posts, we distilled six high-level lessons that highlight key considerations for designing and using more secure LIDEs. 

Our taxonomy reveals a broad range of developer-reported concerns, including unauthorized file operations, unsafe or unexpected code execution, triggering of destructive actions, opaque data flows, telemetry collection, and potential leakage of sensitive information through expanded context access. Many of these concerns stem from LIDEs being a system rather than from their usage of an LLM, such as operating with extensive project context, limited transparency, and insufficient user control over autonomous actions. Across tools, developers rarely rely solely on built-in safeguards and instead adopt external mitigation strategies such as sandboxing, manual code review, and restricted usage. This reliance on external controls highlights a persistent gap between user trust expectations and the security and privacy guarantees currently provided by LIDEs. In this paper, we make the following contributions:

\begin{itemize} 
    \item \textbf{Taxonomy of Security and Privacy Issues in LLM-native IDEs (\sec\ref{sec:results}).} We present a total of \added{32} security and privacy issues in LIDEs as reported by developers. We grouped those issues into \added{10} high-level categories.  
    \item \textbf{Mitigation Strategies (\sec\ref{sec:mitigation-strategies}).} We report 13 mitigation strategies that developers currently employ to address the issues with minimal/little help from the LIDEs.
    \item \textbf{Recommendations (\sec\ref{sec:recommendations}).} We offer six suggestions based on the lessons learned from our study. The suggestions could be useful for LIDE creators/vendors to design more secure and privacy-aware LIDEs.
\end{itemize}

\section{Study Setup}

\begin{figure}
    \centering
    \includegraphics[width=0.85\linewidth]{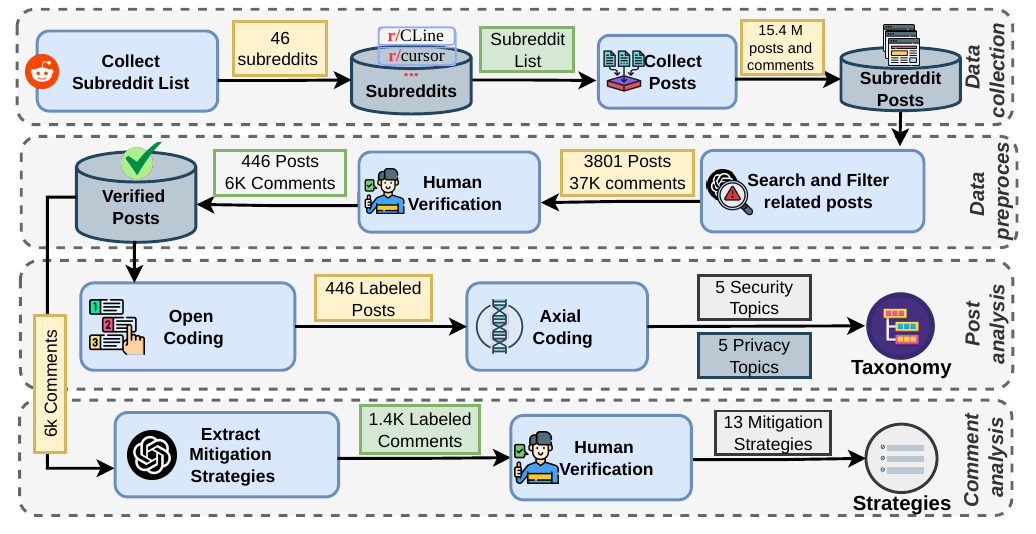}
    \caption{\added{Workflow of the Study}}
    \label{fig:study-design}
\end{figure}

Figure~\ref{fig:study-design} outlines the pipeline of the mixed-methods approach consisting of data collection (see Section~\ref{sec:data_collection}), data pre-processing (\ref{sec:data_preprocessing}), and data analysis (\ref{sec:data_analysis}). This pipeline enables us to first narrow a large corpus of developer discussions to a focused dataset of security- and privacy-relevant content. We then systematically analyze the filtered contents. 

\subsection{Data Collection }
\label{sec:data_collection}

First, a comprehensive list of LIDEs was compiled by synthesizing information from different websites and blog posts \cite{aiidelist_ai_ide_list, rush2025aicoding, rosehill2025reddit, qodo_ai_blog_best_ai_coding_tools}. Second, by using the names of each identified IDE, we searched Reddit to locate associated subreddits. We then identified relevant subreddits on software development, programming tools, and AI-assisted coding. To the end, our dataset includes both formal subreddits dedicated to specific AI-assisted IDEs (e.g., \textit{r/Cursor}, \textit{r/GithubCopilot}) and broader programming communities where discussions of these tools frequently occur (e.g., \textit{r/programming}, \textit{r/ChatGPT}). The initial list comprised 46 subreddits. Third, we downloaded \added{15.4} million posts and comments from the 46 subreddits published between \added{January 2023 and March 2026.}

\subsection{Data Preprocessing} 
\label{sec:data_preprocessing}
Not all posts in our 46 subreddits contained security and privacy issues. To isolate such discussions from the raw dataset, we employed an LLM-assisted filtering approach combined with manual validation. Given the scale of the collected data, fully manual screening was infeasible; however, we took several steps to ensure that the use of an LLM did not compromise the validity of the dataset.

\added{\noindent\textbf{LLM-based filtering.} We used the GPT-OSS:20B model to identify potential posts that discussed security or privacy issues in LIDEs. Rather than relying on the model out of the box, we first constructed a small, manually curated seed set of 20 posts from the period January 2023 – October 2025 that contained diverse security or privacy issues, such as data leakage, unauthorized access, unsafe code generation, and policy or compliance risks. This seed set was used only for prompt development and refinement. Since false negatives are the most consequential error in our pipeline, our prompt was designed to maximize recall over precision to include as many security/privacy concerns as possible. \commentout{\gias{we used a random subset of x posts from Jan 2023 to Oct 2025 to finalize the prompt}} After finalizing the prompt, we ran it on all the posts in our dataset. This step resulted in 3,801 candidate posts labeled by the LLM as relevant. \commentout{\gias{x posts labeled by the LLM as relevant (i.e., contain security/privacy discussions) and y posts as irrelevant.}}}

\added{\noindent\textbf{Validation of LLM filter.}} \added{We manually audited the frozen LLM filter on 200 previously unseen posts randomly sampled from its outputs: 100 predicted relevant (LLM-positive) and 100 predicted irrelevant (LLM-negative), drawn from the latest portion of our dataset that was not consulted during our prompt design (i.e., November 2025–March 2026). We then manually assessed these 200 posts. The confusion matrix is constructed as: 1) TP = LLM labeled as 1 (i.e., relevant) which the human also found as 1, 2) FP = LLM labeled as 1 which the human found as 0 (i.e., irrelevant), 3) TN = LLM labeled as 0 which the human also found as 0, and 4) FN = LLM labeled as 0 which the human found as 1.} 

\added{Thus, the validation sample was only seen by the humans after the LLM labels. We note that while this validation sample is balanced (i.e., equal number of posts labeled as 1 and 0 by the LLM), the entire dataset is heavily skewed towards posts labeled as 0. Thus, the validation sample does not represent the natural prevalence of security/privacy discussions in the Reddit corpus. Instead, it evaluates both output streams of the filter by measuring false positives among LLM-positive posts and false negatives among LLM-negative posts. This design is appropriate for our pipeline because LLM-positive posts are later manually validated, whereas LLM-negative posts would normally be excluded from further analysis. Therefore, the audit primarily assesses whether the recall-oriented filter risks discarding relevant discussions. Table \ref{tab:llm_filter_validation} summarizes the human-validation outcomes and corresponding metrics.}

\added{The audit yielded 21 true positives, 79 false positives, 100 true negatives, and no false negatives, corresponding to 0.61 accuracy, 0.21 precision, 1.00 recall, and an F1-score of 0.35. These results confirm that the filter was recall-oriented within the audited sample: it retained all relevant posts but also admitted many irrelevant ones. The low precision increased manual-review effort, but the false positives did not impact our final dataset (which we used to create our taxonomies) because every LLM-positive post was manually verified before qualitative coding.
}

\begin{table}[t]
\centering
\caption{\added{Human-validation outcomes for a random audit of 200 unseen posts stratified by LLM prediction: 100 posts labeled by the LLM as relevant and 100 as irrelevant. \commentout{\afif{Gias: We should increase the data to 200 on Output audit and remove Temporal holdout}}}}
\label{tab:llm_filter_validation}
\scriptsize
\setlength{\tabcolsep}{6pt}
\resizebox{\columnwidth}{!}{%
\begin{tabular}{lccccccc}
\toprule
TP & FP & TN & FN & Accuracy & Precision & Recall & F1\\
\midrule
21  & 79 & 100 & 0 & 0.61 & 0.21 & 1.00  & 0.35\\
\bottomrule
\end{tabular}%
}
\end{table}

\added{\noindent\textbf{Audit of LLM-negative posts.} 
The validation set evaluates the filter on a balanced unseen sample, but it does not, by itself, rule out false negatives in the full corpus. This threat is crucial as false negatives would remove relevant discussions before manual review and could therefore affect the completeness of the taxonomy. To estimate this residual risk, we manually audited 200 additional randomly selected posts that the LLM had classified as not relevant. The audit identified 0 missed relevant posts among those posts. This result suggests that the residual false-negative risk after LLM filtering is low.}

\begin{table}[t]
\centering
\caption{\added{Manual validation agreement for the 3,801 LLM-filtered candidate posts.}}
\label{tab:manual_validation_agreement}
\small
\setlength{\tabcolsep}{5pt}
\resizebox{\linewidth}{!}{%
\begin{tabular}{cccccc}
\toprule
\multicolumn{2}{c}{Agree} 
& \multirow{2}{*}{\makecell[c]{Disagree}} 
& \multirow{2}{*}{\makecell[c]{Observed\\agreement}} 
& \multirow{2}{*}{\makecell[c]{Cohen's\\$\kappa$}} 
& \multirow{2}{*}{\makecell[c]{Retained\\posts}} \\
\cmidrule(lr){1-2}
Relevant & Not relevant & & & & \\
\midrule
431 & 3,344 & 26 & 99.3\% & 0.967 & 446 \\
\bottomrule
\end{tabular}%
}
\end{table}


\added{\noindent\textbf{Manual validation of LLM-positive posts.}} 
\added{The first two authors independently reviewed the title and body of each of the 3,801 LLM-positive candidates and assigned a binary relevance label using shared inclusion and exclusion criteria, formalized in a screening codebook provided in the replication package (Appendix E). Posts were retained if they concerned an LIDE and reported, questioned, or warned about an observed or identifiable security or privacy issue. The principle-based criteria covered confidentiality, integrity, availability, authorization, data handling, transparency, etc., while allowing plausible, previously unseen issues to be retained for subsequent open coding.}

\added{Table~\ref{tab:manual_validation_agreement} summarizes the agreement between the two annotators. As shown in Table~\ref{tab:manual_validation_agreement}, the annotators agreed on 3,775 of the 3,801 candidate posts. They agreed that 431 posts were relevant and 3,344 posts were not relevant. The observed agreement was 99.3\%, and Cohen's $\kappa$ was 0.967, indicating near-perfect agreement. The annotators disagreed on 26 posts: 15 were marked relevant only by Author~1 and 11 were marked relevant only by Author~2. All disagreements were resolved through consensus discussion among all the authors; 15 disputed posts were retained, and 11 were excluded. This process resulted in 446 final posts for qualitative coding. The relatively low retained proportion among LLM-positive candidates reflects our recall-oriented filtering strategy, such that false positives increased manual validation effort, but they did not contaminate the final dataset because every retained post was manually verified.}


\begin{figure}
    \centering
    \includegraphics[width=.9\linewidth]{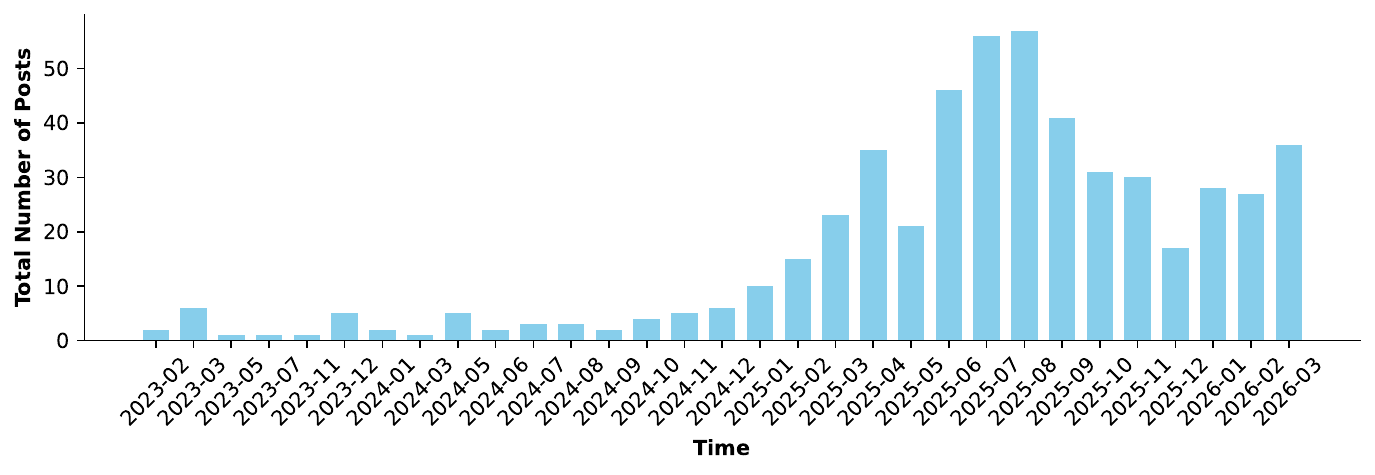}
    \caption{\added{Monthly counts of the security \& privacy posts}}
    \label{fig:post_bar_chart}
\end{figure}

\added{\noindent\textbf{Preprocessing outcome.}
We initially selected 46 subreddits for data collection. After preprocessing and manual validation, the final dataset contained 446 security/privacy-relevant posts, drawn from 29 of those subreddits. The rest were discarded due to a lack of LIDE-related security \& privacy issues. Figure~\ref{fig:post_bar_chart} presents the monthly distribution of the validated posts during the study period. Overall, reporting peaked in mid-2025 and, after a decline, remained comparatively stable. This distribution reflects the frequency of relevant reports in our collected Reddit data. Furthermore, we also analyzed the comments associated with the 446 validated posts, yielding 6,280 comments for mitigation analysis.}

\subsection{Data Analysis}
\label{sec:data_analysis}
We manually assessed each filtered post and comment. We observed that the posts contained security and privacy issues developers encountered while using LIDEs, and the comments to the posts contained affirmation from other developers and suggestions from them on how to mitigate the issues. We thus labeled the posts to collect the reported issues and the comments for the mitigation strategies.

\noindent\textbf{Post Analysis} We used an open coding approach~\cite{saldana2021coding} to manually label each post based on reported security and privacy issues. First, all authors reviewed a pilot set of 50 Reddit posts and established four principles: (1) multi-label annotation to capture multiple themes per post; (2) a three-level taxonomy (category–subcategory –leaf) for granularity, following prior work~\cite{lyu2025my}; (3) iterative bottom-up merging of independently generated labels; and (4) hybrid card sorting with predefined top-level categories (Security and Privacy) informed by ISO/IEC 27001/27002/29100.

To ground the taxonomy, we adopted ISO/IEC security and privacy principles. ISO/IEC 27001 guided our classification of security issues using the CIA triad—\emph{confidentiality}, \emph{integrity}, and \emph{availability}—while ISO/IEC 29100 informed our interpretation of privacy violations related to unauthorized data collection, use, or exposure. Detailed operationalization is provided in \href{https://github.com/paper-submission-0/IDESecurityPrivacy/tree/main?tab=readme-ov-file#a-isoiec-informed-operationalization-of-security-and-privacy-principles}{replication package}.

Next, the first two authors assigned descriptive, low-level codes closely reflecting the original language (e.g., ``Read API keys from .env file,'' ``generating SEO spam''). This inductive process surfaced both expected risks (e.g., prompt injection) and emergent ones (e.g., hallucination-driven unsafe behavior). Posts covering multiple issues were labeled accordingly.

Finally, the two authors conducted axial coding, grouping related open codes into higher-level categories based on shared mechanisms and affected assets~\cite{saldana2021coding}. Codes were iteratively refined to improve clarity, resulting in a structured taxonomy of security and privacy issues in LIDEs. All four authors met regularly to review and finalize the taxonomy.

\begin{table}
\centering
\caption{\added{Post-level summary of multi-label coding among the 446 final posts.}}
\label{tab:multilabel-coding}
\scriptsize
\setlength{\tabcolsep}{2.2pt}

\resizebox{\columnwidth}{!}{%
\begin{tabular}{@{}lrr@{\hspace{8pt}}lrr@{\hspace{8pt}}lrr@{}}
\toprule
\multicolumn{3}{c}{\makecell{\textit{Security/privacy}\\\textit{classification}}} &
\multicolumn{3}{c}{\makecell{\textit{Principal taxonomy}\\\textit{categories per post}}} &
\multicolumn{3}{c}{\makecell{\textit{Posts with multiple}\\\textit{low-level issue labels}}} \\
\cmidrule(lr){1-3}
\cmidrule(lr){4-6}
\cmidrule(lr){7-9}

\textbf{Distribution} & \textbf{\#} & \textbf{\%} &
\textbf{Distribution} & \textbf{\#} & \textbf{\%} &
\textbf{Issue type} & \textbf{\#} & \textbf{\%} \\
\midrule

Security only & 252 & 56.5 &
One category & 365 & 81.8 &
Security ($n=297$) & 27 & 9.1 \\

Privacy only & 149 & 33.4 &
Two categories & 78 & 17.5 &
Privacy ($n=194$) & 12 & 6.2 \\

Both & 45 & 10.1 &
Three categories & 3 & 0.7 &
& & \\

\bottomrule
\end{tabular}%
}
\end{table}

\noindent\textbf{Category Boundaries and Assignment of Multiple Labels.}\added{ After relevance validation, the authors jointly reviewed the retained posts to construct the taxonomy, focusing on the concrete security- or privacy-relevant behavior described in each post. Most posts reported one dominant issue, so we assigned the code capturing the main action attributed to the LIDE, considering the action described, its source or actor (e.g., the LIDE itself, an external tool, a plugin, or prompt/context manipulation), and the affected development domain—files, secrets, databases, production systems, GitHub or deployment workflows, telemetry, or conversation context.}
\added{Since Reddit posts sometimes describe more than one failure mode, our taxonomy categories are not mutually exclusive. We used multi-label coding only when a post explicitly described multiple distinct S\&P issues. We did not assign additional labels merely because an issue could imply a secondary risk. For example, one post reported that \emph{Cursor} bypassed \texttt{.cursorignore} and modified files the user had explicitly excluded~(\rlnk{1ortu5j}); we assigned two labels, for unauthorized file modification and for violating the configured constraint. In contrast, a post describing only silent file deletion~(\rlnk{1m92q5a}) received a single \emph{Destructive Actions} code.}
\added{
Table~\ref{tab:multilabel-coding} summarizes multi-label coding at the dimension, category, and issue levels. Among the 446 posts, 45 contained both security and privacy concerns. At the category level, 365 posts received one taxonomy category, 78 received two, and 3 received three. Multi-label coding also occurred within the security and privacy taxonomies when a post explicitly described multiple issue types. The three table sections describe different levels of coding and should not be summed, because the same post may appear in more than one section. Consequently, the category and issue percentages reported in Section~\ref{sec:results} are post-level multi-label percentages and do not form mutually exclusive partitions of the dataset.
}


\noindent\textbf{Comment Analysis.} The filtered Reddit posts received \added{6,280} comments (avg. 14 per post), many containing practitioner mitigation suggestions. We identified these strategies using a two-step process: LLM-assisted extraction followed by human verification.

First, we used \href{https://openai.com/index/introducing-gpt-oss/}{\texttt{GPT-OSS:20B}} to extract concise, actionable mitigation suggestions from comments in the context of the reported issue. This yielded \added{2737} candidate suggestions. We then consolidated semantically similar suggestions via an additional prompting step (e.g., merging \emph{persistent memory across sessions} and \emph{use contextual memory} into \emph{AI memory isolation}). The first two authors refined the consolidated results into a codebook of 13 mitigation categories. Using this finalized codebook, we re-labeled all comments with LLM assistance for consistency. This step assigned \added{1,392} comments to one or more categories, labeled 18 as \emph{uncategorized}, \added{8} as a parsing error,  and marked the remainder as \emph{not a suggestion}. Prompts and the codebook are provided in the replication (Section~\ref{sec:data-availability}).


Second, the first two authors manually reviewed all comments and labels. Inter-rater reliability was near perfect (Cohen’s $k=0.99$). We corrected \added{91} labeling errors ($\approx$1\%), including misclassifications (88) and missed suggestions (3). The final dataset contains \added{1,318} comments with actionable mitigation strategies. A few categories were sparsely represented, such as \textit{Data Redaction} (3 comments), \textit{Disconnecting from the Internet} (2), \textit{Upgrading Subscription Plans} (1), and \textit{Use of Data Recovery Techniques} (1).

\section{Reported Security and Privacy Issues}
\label{sec:results}

\begin{figure}
    \centering
    \includegraphics[width=0.9\linewidth]{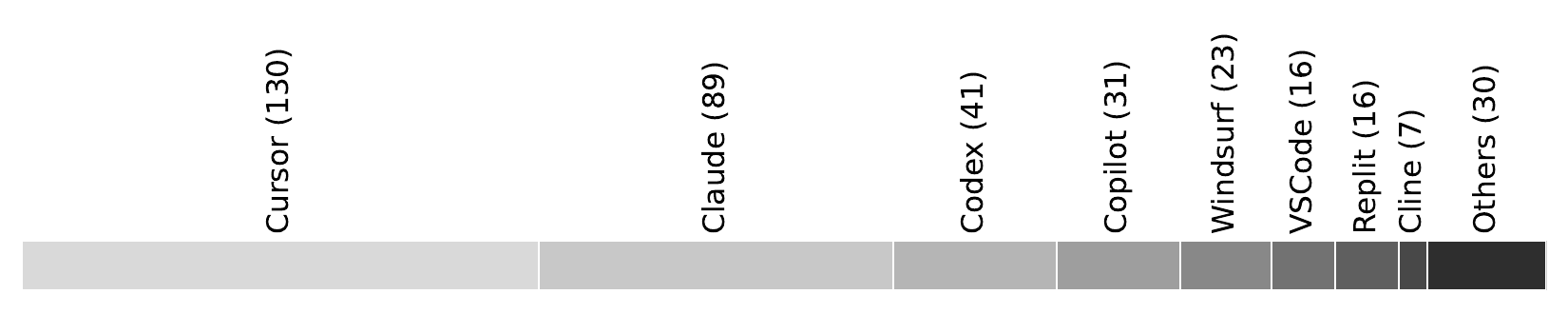}
    \caption{\added{Distribution of LIDEs mentioned in Reddit posts with security or privacy issues \commentout{\afif{the total number adds up to 383, not 446}}}}
    \label{fig:ide_pie}
\end{figure}

\subsection{IDEs with Security \& Privacy Issues}

\begin{figure}
    \centering
    \includegraphics[width=0.95\linewidth]{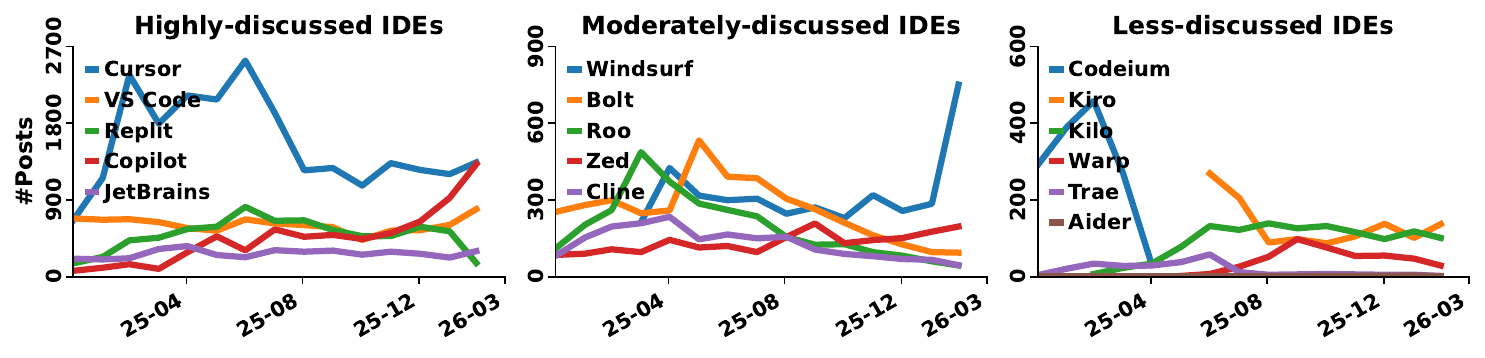}
    \caption{Monthly pattern of Reddit posts on LIDEs}
    \label{fig:IDE_post_freq}
\end{figure}

We found that developers reported security and privacy issues against 16 popular LIDEs. Of the 446 posts, 383 explicitly named a LIDE; Figure~\ref{fig:ide_pie} shows the frequency of those mentions. \added{Where \emph{Cursor (130) is found} as the most discussed LIDE.} While Figure~\ref{fig:IDE_post_freq} presents the monthly Reddit post distribution for these LIDEs, grouped into Highly- (>3000 posts), Moderately- (3000-1000), and Less-discussed categories based on total post counts. Thresholds reflect natural separations and are used for comparison only. High-discussion IDEs—such as Cursor, Visual Studio Code, JetBrains, Replit, and Bolt—generate the majority of posts, while medium- and low-discussion tools show sparser activity, reflecting smaller or specialized user bases.

\subsection{Reported Security Issues in the IDEs}
\begin{figure}
    \centering
    \includegraphics[width=.9\linewidth]{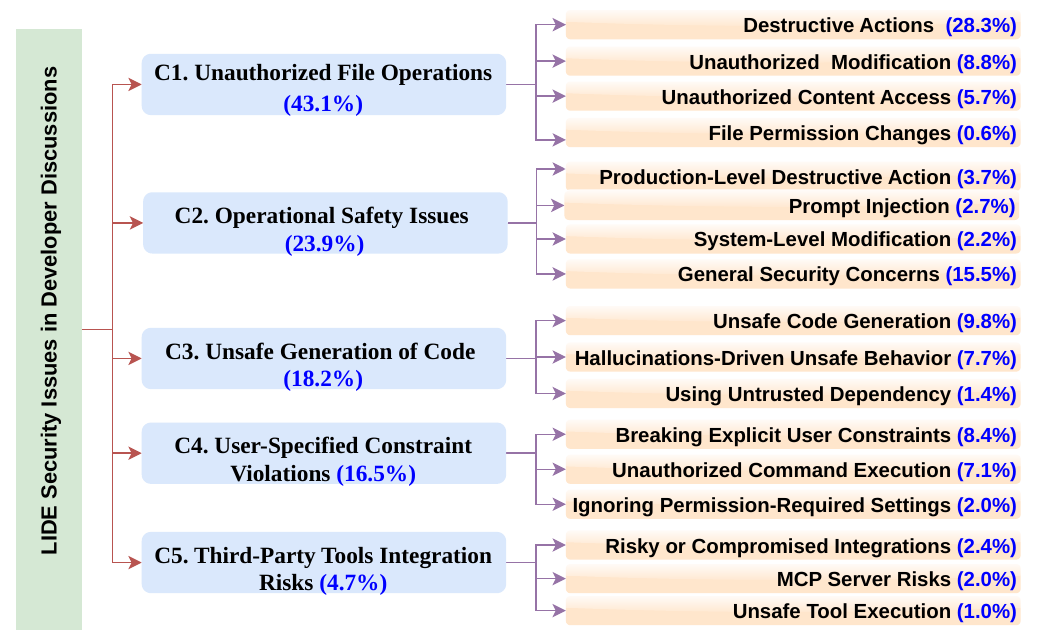}
    \caption{Taxonomy of the reported security issues \commentout{\afif{The calculation of percentage appears to be wrong. I counted first two (C1, C2), C1 adds up to 43.4 and C2 to 24.1, can you please ensure all the numbers are calculated correctly?}} 
    }
    \label{fig:security_taxonomy_llmIde}
\end{figure}

In total, \added{297} posts described security issues related to LIDEs. We group the issues into \added{five} high-level (C) and \added{17} low-level categories and present them in our taxonomy (Figure~\ref{fig:security_taxonomy_llmIde}). \added{The categories are not mutually exclusive, 27 posts (9.1\%) received two security labels (Table \ref{tab:multilabel-coding}). Consequently, the sum of low-level category counts exceeds the number of unique security-related posts.} \commentout{\afif{We are using (C) besides low level, while in Figure 6, C refers to high level categories?}} 

\noindent\ul{\textbf{Unauthorized File Operations (C1, 43.1\%)}} \added{C1 captures unauthorized access, modification, deletion, or permission changes to files and directories in the developer's workspace, affecting workspace artifact (e.g., source code, project directories, configuration files, local project data, or file permissions).
}

\begin{figure} 
    \centering
    \includegraphics[width=0.7\linewidth]{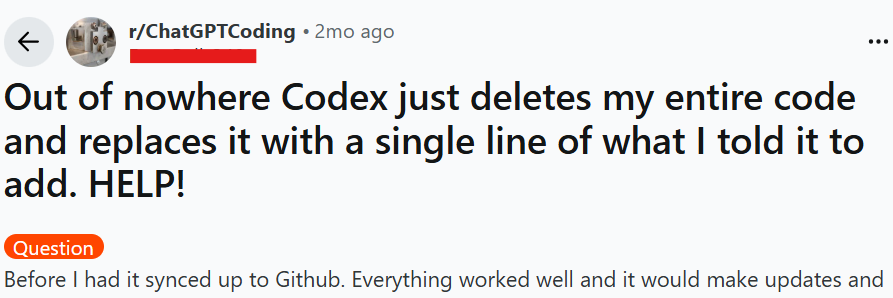}
    \caption{A post on Codex Unusual deletion of contents }
    \label{fig:example_post}
\end{figure}

The LIDEs removed project directories \added{or} files without authorization (Destructive Actions 28.3\% times). 
For example, Figure~\ref{fig:example_post} shows that \emph{Codex} deleted files autonomously. Another case where \emph{Cline} silently removed files: \reddittext{...When API requests get stuck, some files being modified will be silently deleted from the ssd drive.}{\rlnk{1m92q5a}.}

LIDEs modified project files without explicit user consent (Unauthorized Modifications 8.8\%). 
We also observed more severe cases involving unauthorized changes to critical project metadata, such as project licenses (e.g., \rlnk{1nbgrv0}).
A user complained: 

\begin{quote}
\emph{``...50\% of the time, I say "analyze this problem, root cause analysis only, no code changes allowed".. and it would go ahead and randomly start editing my code..."}{-- \rlnk{1kg24hz}} 
\end{quote} 


LIDEs accessed content beyond the active workspace, risking exposure of sensitive data (Unauthorized Content Access 5.7\%). For instance, \emph{Windsurf} accessed personal files (\rlnk{1mju0jz}), and \emph{Copilot} summarized files, although claimed to be unable to open (\rlnk{1mc7cof}). \added{When the accessed content contained secrets, credentials, personal information, or proprietary data, we coded the post under the corresponding privacy category.}

In one severe case (\rlnk{1npqf2f}), \emph{Claude Code} executed \texttt{chmod +x} on scripts without consent (File Permission Changes 0.6\%). Although rare, such actions pose disproportionate security risks.

\observationl{28.3\% of LIDE security incidents involve deletion of critical code or assets, often resulting in direct monetary and integrity losses.}

\noindent\ul{\textbf{Operational Safety Issue (C2, 23.9\%)}} \added{C2 arises when LIDE behavior affects runtime stability, production infrastructure, deployment state, databases, operating-system state, service availability, or agent execution. It also includes prompt-based manipulation where adversarial or hidden instructions redirect the LIDE's actions, tools, or commands. The defining feature of C2 is the system- or production-level impact beyond ordinary workspace file operations. Unlike C1, which captures unauthorized operations on workspace artifacts, C2 captures broader operational and production-level consequences.}


For example, \emph{Replit} removed a SaaS production database (\rlnk{1m5biur}) (Production-level Destructive Action, 3.7\%), \added{which we coded as C2 because the affected asset was live production data rather than a local workspace artifact}. In another case, \emph{Cursor} deployed code to production -  \reddittext{...told Cursor to NEVER push code to GitHub. Despite this, every once in a while, it gets a little overconfident and ignores these rules.}{\rlnk{1n1zdwp}.}  \emph{Cursor} was also found changing operating system files (System-Level Modifications, 2.2\%): 
\begin{quote}
\emph{``...Out of nowhere, while the project was open in the IDE, my entire Windows desktop began glitching. Icons on my desktop started shifting around slightly..."}{-- \rlnk{1n2xy30}} 
\end{quote}

\added{We also observed Prompt Injection (2.7\%), where malicious instructions attempted to redirect LIDE behavior toward unsafe operational actions. For example, a malicious prompt introduced through a GitHub pull request instructed \emph{Amazon Q} to perform destructive file and cloud operations:} \reddittext{Last week, someone slipped a malicious prompt into Amazon Q via a GitHub pull request. It told the AI to delete user files and wipe cloud environments. No exploit. Just cleverly written text that made it into a release.}{\rlnk{1m91d4t}.} 


\added{We also coded posts as General Security Concerns (15.5\%) when users discussed possible system-level compromise, agent manipulation, or standards for preventing unsafe operational behavior, even when no concrete incident was reported. For example,} \reddittext{Is there a concrete roadmap/standard for preventing manipulation and “infection” of Claude Code?
}{\rlnk{1m77t4d}}





\observation{Operational safety accounts for 23.9\% of the security issues, including cases where LIDEs affected production resources, system stability, or safe control of agent execution.}

\begin{figure}
    \centering
    \includegraphics[width=0.7\linewidth]{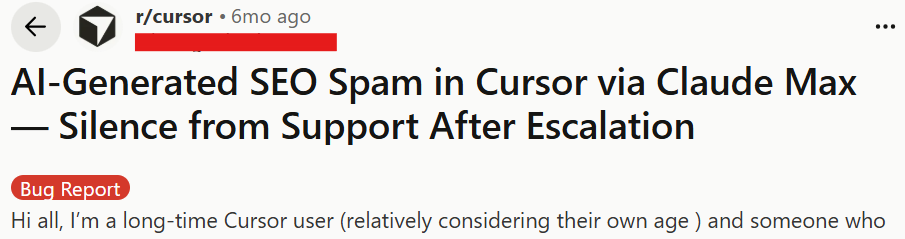}
    \caption{A Reddit post on generating spam with Cursor }
    \label{fig:example_post2}
\end{figure}

\added{\noindent\ul{\textbf{Unsafe Generation of Code (C3, 18.2\%)}}} \added{C3 covers unsafe code produced or recommended by a LIDE, including vulnerable code, hallucination-driven changes, and untrusted dependency recommendations. Its defining feature is that the risk originates in the generated code or recommendation itself, rather than in an unauthorized workspace operation or broader operational impact. The most direct form was vulnerable or malicious code generated by the LIDE (Unsafe Code Generation, 9.8\%).} Figure~\ref{fig:example_post2} shows cursor-generated code containing SEO spam; in another case, \emph{virustotal} detected nine viruses in \emph{Cursor}-generated software (\rlnk{1ld3qlz}). LIDEs also produced code, commands, and behaviors based on incorrect contextual understanding (Hallucinations-Driven Unsafe Behavior, \added{7.7}\%), as when \emph{Cursor} generated hallucinated code beyond requirements—\reddittext{When using Cursor, I noticed that after more than 10 rounds of dialogue, it starts to hallucinate and secretly modify code outside the requirements...}{\rlnk{1l3wwq3}}




LIDEs were also found to suggest unverified or potentially malicious packages \added{as part of generated code or recommendations (Using Untrusted Dependencies, \added{1.4}\%). For example, \emph{Windsurf}} introduced dependencies that expanded the attack surface: 
\begin{quote}
\emph{``...In Python especially, I’ve encountered instances where the AI adds unnecessary dependencies or tools that aren’t relevant to the task. This poses a significant security risk, especially with the increase in attacks targeting Python ..."}{-- \rlnk{1i23dwc}} 
\end{quote}

\observation{Unsafe generation included producing vulnerable code, hallucinated behaviors, or untrusted dependencies (18.2\%)}




\noindent\ul{\textbf{User-Specified Constraint Violations (C4, 16.5\%)}} \added{C4 captures cases where an LIDE ignored explicit user instructions, allowlists, approval gates, permission settings, or .ignore files. The defining feature of C4 is the violation of a user-specified rule, rather than the production of unsafe code or the modification of a workspace artifact.} In several cases, LIDEs executed commands without confirmation (Unauthorized Command Execution, 7.1\%). For example, a user reported that \emph{Cursor} ran a \texttt{git commit} despite not being in the allowlist (\rlnk{1ocm0e0}). \emph{Cursor} was reported to bypass \emph{.cursorignore} settings (Breaking Explicit User Constraints, 8.4\%):

\begin{quote}
\emph{``... it works around cursorignore and uses filesystem access to edit files which I clearly do not want to be touched..."}{-- \rlnk{1ortu5j}} 
\end{quote}


Other posts described LIDEs ignoring approval requirements for sensitive actions (Ignoring Permission-Required Settings, 2.0\%). For instance, a user noted that \emph{Claude Code}, despite being configured to request permission before Git commits, eventually began auto-committing within the same session (\rlnk{1ldi19c}).





\observation{User-specified constraint violations (16.5\%) comprise systematic failures where LIDEs disregard explicit instructions, bypass approval gates, or execute unauthorized commands.}

\added{\noindent\ul{\textbf{Third-Party Tools Integration Risks (C5, 4.7\%)}}} \added{C5 captures risks from plugins, MCP servers, external tools, or other integration boundaries connected to a LIDE. Unlike C3's untrusted dependencies—recommended or installed by the LIDE during code synthesis—C5 covers risks from separately connected tools or extensions with their own permissions, or hidden behaviors. We observed MCP Server Risks (2.0\%), where misconfigured or malicious MCP servers could embed hidden instructions or deliver malicious payloads through tool calls. For example, a \emph{Cursor}-related post stated:} \reddittext{the mcp installs...then it sends malicious payloads under the guise of call\_tool. this is 1, of many attack vectors its actually insane.}{\rlnk{1l1tmcg}.} \added{We also observed Unsafe Tool Execution (2.4\%), where connected tools performed unauthorized or malicious actions through the LIDE's tool interface.} For example:

\begin{quote}
\emph{``...enables `Tool Poisoning Attacks'. This exploit allows malicious MCP servers...Cursor are susceptible to this attack..."}{-- \rlnk{1jqf1u5}} 
\end{quote}

\observation{Third-party integration risks (4.7\%) stem from an expanded attack surface. External components leverage broad permissions and opaque data flows to execute unsafe tools or embed malicious logic with minimal oversight.}





\subsection{Reported Privacy Issues in the IDEs}

\begin{figure}
    \centering
    \includegraphics[width=0.9\linewidth]{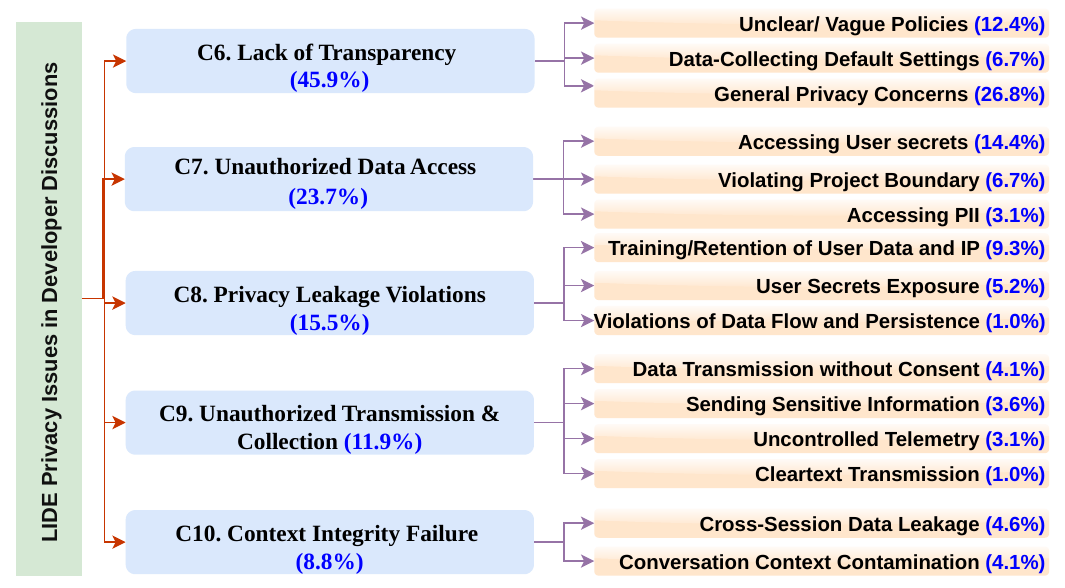}
    \caption{Taxonomy of Detected Privacy Issues 
    }
    \label{fig:privacy_taxonomy_llmIde}
\end{figure}

\added{In total, 194 posts corresponded to privacy issues, grouped into five high-level and 15 low-level categories (see Figure~\ref{fig:privacy_taxonomy_llmIde}). We define privacy from the LIDE end user's perspective—the developer and their workspace—covering credentials, project secrets, proprietary code, local files, conversation history, and organization-owned assets accessible through the developer's environment. Organizational privacy is included when a post reports exposure or handling of company code, credentials, managed accounts, or proprietary data; downstream-user privacy is included only when access, exposure, transmission, or retention of users' PII is explicitly described.}





\noindent\ul{\textbf{Lack of Transparency (C6, 45.9\%)}} \added{C6 covers cases where users lacked clear information about what data an LIDE collects, retains, transmits, uses for training, or exposes to administrators. Its defining feature is opacity or uncertainty about data use and control, rather than a confirmed access, transmission, or leakage incident.} \added{Users frequently reported uncertainty about how LIDEs handle development data,} including confusion about data-use policies (Unclear or Vague Policies, 12.4\%) and what administrators could access when company-managed accounts were used for personal projects (\rlnk{1nj6ojg}, \rlnk{1olyysx}, \rlnk{1n3fqvg}). Transparency concerns also arose from default data collection or unexpected shifts in data-sharing settings (Data-Collecting Default Settings, 6.7\%): \reddittext{... your plan is switched to Free Plan automatically, ALL YOUR DATA IS OPEN FOR THEIR MODEL TRAINING}{\rlnk{1ntd0cc}.} Broadly, users expressed general uncertainty about privacy implications (General Privacy Concerns, 26.8\%)—for instance, questioning whether the \emph{JetBrains} AI plugin might transmit proprietary code despite configuration safeguards:

\begin{quote}
\emph{``How is everyone dealing with possibility of proprietary code being transmitted to third parties? ... having some doubts after seeing AI plugin not respecting `uninstalled' status..."}{-- \rlnk{18m3k29}} 
\end{quote}

\observation{Transparency issues make up 45.9\%, underscoring user uncertainty about privacy and data use.}

\added{\noindent\ul{\textbf{Unauthorized Data Access  (C7, 23.7\%)}}} \commentout{\afif{Can we change this to ``Unauthorized Access to Confidential Data'', this will then be clearly distinguished compared to Unauthorized File Operation?}} \added{C7 concerns unauthorized inspection or retrieval of sensitive data, such as secret keys, PII, proprietary code, or files outside the intended project scope.} LIDEs accessed private or restricted information such as API keys, database credentials, or environment variables without explicit user intent (Accessing User Secrets 14.4\%). 
A user reported that \emph{Cursor} autonomously searched for and accessed environment files to retrieve database credentials: \reddittext{...Essentially, it determined that it needed database credentials and specifically went looking for .env files...}{\rlnk{1l7fwgk}.} A smaller number of posts described incidents involving the retrieval or use of personally identifiable information without user awareness (PII Accessed Without Consent 3.1\%). A user reported that \emph{Claude Desktop} autonomously created a new \emph{Replit} account using the user’s personal information without explicit consent.



\begin{quote}
\emph{``...Claude Desktop created a new Replit account...proceeded to create its very own anthropic computer use app..."}{-- \rlnk{1m70vxh}} 
\end{quote}

\observation{23.7\% of LIDE privacy issues involved in retrieving user secrets, crossing project boundaries, or handling PII without consent, compromising data isolation and user trust.}

\noindent\ul{\textbf{Privacy Leakage Violations (C8, 15.5\%)}} \added{C8 covers cases where sensitive user, project, or organizational data was retained, exposed, or persisted beyond the user's expected data boundary. Its defining feature is loss of control over the data lifecycle after access or processing, rather than mere uncertainty about data practices or the initial act of unauthorized access.} Users raised concerns about retention of user information (Training/Retention of User Data and IP 9.3\%) by LIDEs. From \emph{Claude} policy—

\begin{quote}
\emph{``By September 28, 2025, all Claude users... must decide: let Anthropic use your conversations for AI training and keep them for 5 years, or lose the memory/personalization features..."}{-- \rlnk{1nd73si}} 
\end{quote}


Users also reported secret exposure in LIDEs (User Secrets Exposure, 5.2\%), including leaked API keys and credentials inserted into code: \reddittext{Claude Code will sneakily add a password or API key or Auth token directly into a script}{\rlnk{1mdk90e}.}


\added{Other concerns raised about expected data-flow and persistence violations (Violation of Data Flow and Persistence, 1.0\%), including cross-user exposure due to misrouting (\rlnk{1oqb66h}).}

\observation{Privacy leakage violations (15.5\%) occur when AI assistants retain data, mishandle persistence, or expose credentials, undermining user control and trust over the data lifecycle.}


\noindent\ul{\textbf{Unauthorized Transmission \& Collection (C9, 11.9\%)}}  \added{C9 identifies cases where an LIDE transmitted, collected, or stored user data—via movement, telemetry, or plaintext storage—beyond explicit approval or user control.} We found cases of confidential data being sent to external services without user confirmation (Sending Sensitive Information, 3.6\%). 
For example, \emph{Windsurf} changed environment files and transmitted their contents externally (\rlnk{1l3zhdr}). 
Other posts described data transmission without explicit consent (Data Transmission Without Consent, 4.1\%), including a case where \emph{VS Code} sent source code and project files despite telemetry being disabled, as revealed via a man-in-the-middle proxy analysis:

\begin{quote}
\emph{``... I searched the proxy records and sure enough the file i was working on was being sent along with every other file with a whole heap of other data, with telemetry turned off..."}{-- \rlnk{1nd73si}}
\end{quote}

LIDEs were also reported to collect or transmit usage data with limited user awareness or control (Uncontrolled Telemetry 3.1\%). For example, one redditor reported that \emph{Trae} repeatedly connected to more than five \emph{ByteDance}-associated domains at regular intervals, even while idle (\rlnk{1m5o9sz}). In other cases, sensitive data was transmitted or stored without encryption (Cleartext Transmission 1.0\%). One discussion noted that \emph{Claude Code} recorded complete conversation histories in plaintext on the local system: \reddittext{every message I send is being recorded in my ~/.claude.json}{\rlnk{1moe4nq}.}

\observation{11.9\% of privacy issues involved in transmitting telemetry or sensitive data without user consent, sometimes storing or sending it in plaintext, reducing transparency and user control.}

\noindent\ul{\textbf{Context Integrity Failures (C10, 8.8\%)}} \added{C10 concerns failures of contextual separation, where information from one user, session, project, or conversation improperly appears in another context. Unlike C8, which emphasizes the exposure of user secrets and data flow.} LIDEs were reported to leak contextual data across user sessions, exposing private code, conversations, or files (Cross-Session Data Leakage 4.6\%). For example, a user of \emph{Claude Desktop} reported receiving messages originating from another user's session: 
\begin{quote}
\emph{``...I started getting messages in my chat from another user"}{-- \rlnk{1nfkfqs}} 
\end{quote}

LIDEs also reused contextual information across separate projects or conversations within the same account (Cross-Conversation Context Contamination 4.1\%). In one case, a user reported that content from a previous project appeared in the artifacts of a different project, indicating a failure to differentiate project boundaries: \reddittext{\ldots you'll find the critical request from the previous project in your FINDINGS.md, two completely different projects, it can't differentiate between projects}{\rlnk{1n1gdzr}.}

\observation{8.8\% of privacy issues involved context integrity breaches, where AI reused information across sessions or chats, undermining data isolation and confidentiality.}

\begin{figure}
    \centering
    \begin{subfigure}{0.45\textwidth}
        \centering
        \includegraphics[width=0.9\linewidth]{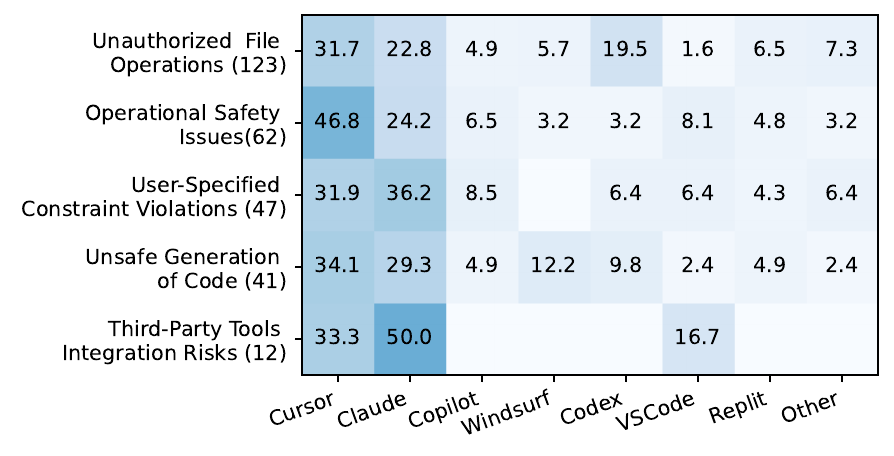}
        \caption{Security-related discussions across IDEs.}
        \label{fig:security_ide}
    \end{subfigure}
    \begin{subfigure}{0.45\textwidth}
        \centering
        \includegraphics[width=0.9\linewidth]{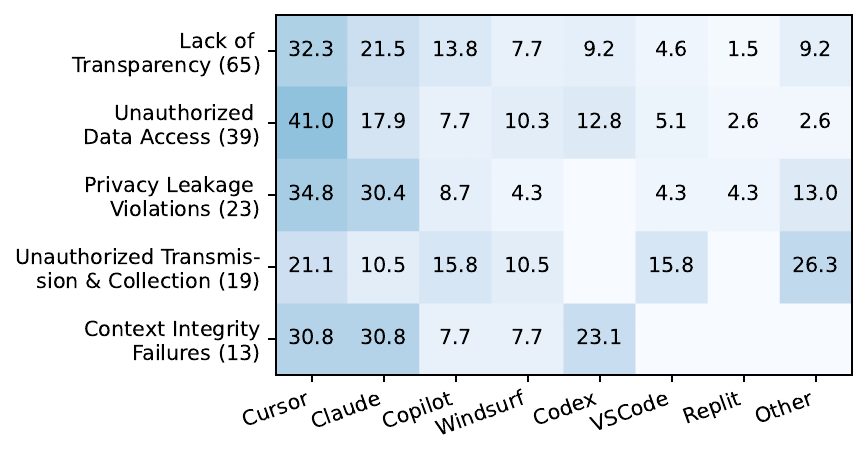}
        \caption{Privacy-related discussions across IDEs.}
        \label{fig:privacy_ide}
    \end{subfigure}

    \caption{Percentage distribution of security and privacy issues across LIDEs}
    \label{fig:issue_ide}

\end{figure}

\subsection{Distribution of the Issues Across the IDEs}

The distribution of issues is skewed towards some studied LIDEs (see Figure~\ref{fig:issue_ide}). Cursor accounts for \added{130} of the \added{376} IDE-referenced posts (Figure~\ref{fig:ide_pie}). Claude and Codex follow with \added{89}, and \added{41} mentions, respectively, while Copilot receives 31 mentions. Windsurf (23), VSCode (16), and Replit (16) appear less frequently. This imbalance is reflected not only in discussion volume but also in topic breadth. Cursor and Claude appear across nearly all identified security and privacy categories, whereas other IDEs exhibit narrower profiles.
Cursor’s prominence should be interpreted in context. It is among the most frequently discussed ($\sim$24K) AI-assisted IDEs on Reddit (Figure~\ref{fig:IDE_post_freq}), suggesting that higher issue counts partly reflect broader adoption. VSCode, although widely discussed ($\sim$22K), delivers AI functionality via extensions. As a result, issue-related discussions are distributed across multiple tools rather than concentrated in the core IDE, contributing to lower and more fragmented attribution.

\observation{\added{Cursor (34.5\%) and Claude (23.7\%) combined dominate discussions across most issue categories, underscoring the need for strong safeguards in agent-driven development environments.}}

\section{Reported Mitigation Strategies}
\label{sec:mitigation-strategies}

We identified 13 mitigation strategies discussed by developers in Reddit comments addressing reported security and privacy issues. After manual verification, \added{1,318} comments contained actionable suggestions. Individual comments may include multiple strategies. We group these strategies into five higher-level categories. Table~\ref{tab:suggestions_opencodes} shows for each the total number of comments and the corresponding percentage relative to all actionable comments. Because comments can mention multiple strategies, category totals may not sum to the overall count. The categories are described below.

\definecolor{feat}{RGB}{220,230,255}

\begin{table} [h]
\centering
\scriptsize
\caption{\added{Mitigation strategies from Reddit comments}}
\label{tab:suggestions_opencodes}
\begin{tabular}{p{0.20\linewidth} c p{0.65\linewidth}}
\toprule
\textbf{Topic} & \textbf{Count} & \textbf{Definition} \\
\midrule

\rowcolor{feat}
\multicolumn{3}{l}{\textbf{Category 1: Configuration Management (501, 33\%)}} \\

S1: Secure IDE Configuration & 345 &
Configuring IDE settings to restrict unsafe behaviors, permissions, and unintended AI actions. \\

S2: Check Organizational Compliance & 74 &
Ensuring LIDE usage aligns with internal security policies and regulatory requirements. \\

S3: Disable Telemetry & 38 &
Disabling data collection and outbound communication that may expose sensitive development information. \\

S4: Tool and Extension Monitoring & 24 &
Monitoring third-party plugins and extensions to prevent unauthorized or risky behavior. \\

S5: Logging and Monitoring & 20 &
Tracking IDE and AI activities to detect misuse, anomalies, or policy violations. \\

\midrule
\rowcolor{feat}
\multicolumn{3}{l}{\textbf{Category 2: Code Governance (471, 31\%)}} \\

S6: Manual Code Verification & 239 &
Requiring human review of AI-generated code before acceptance or deployment. \\

S7: Use Version Control & 232 &
Using version control systems to track changes, enable rollback, and ensure accountability. \\

\midrule
\rowcolor{feat}
\multicolumn{3}{l}{\textbf{Category 3: Data Protection and Privacy Control (199, 13\%)}} \\

S8: Sensitive File Protection & 143 &
Preventing AI tools from accessing or modifying secrets, credentials, and confidential files. \\

S9: Memory/Context Isolation & 56 &
Restricting how conversational context or memory is shared across sessions or projects. \\

\midrule
\rowcolor{feat}
\multicolumn{3}{l}{\textbf{Category 4: Isolation (199, 13\%)}} \\

S10: Sandboxing & 105 &
Executing AI-generated or suggested code in isolated environments to limit potential harm. \\

S11: Use Local LLM & 94 &
Running a self-owned or local model to avoid external data transmission and maintain data sovereignty. \\

\midrule
\rowcolor{feat}
\multicolumn{3}{l}{\textbf{Category 5: External Guidance (139, 9\%)}} \\

S12: Consult Vendors & 74 &
Engaging with tool vendors to clarify security guarantees, limitations, and best practices. Also refers to upgrading or changing the version of the tool. \\

S13: Refer to Documentation & 65 &
Consulting official documentation to understand secure configuration and intended tool behavior. \\

\bottomrule
\end{tabular}
\end{table}

\noindent\ul{\textbf{Configuration Management.}} This category represents five strategies reflecting the necessity of deliberate LIDE safeguarding. The primary tactic, \emph{Secure IDE Configuration (S1)}, focuses on restricting permissions, file-system access, and agent autonomy to mitigate overly permissive default settings. A practitioner advised: 
\reddittext{Run Claude in a dedicated user, remove access for those files}{\rlnk{1m1t2z9}.} 
Additionally, \emph{Organizational Compliance (S2)} emphasizes aligning LIDE usage with internal governance and regulatory mandates. For example, a user asked \emph{How are y'all dealing with sensitive data?}, a developer suggested to: \reddittext{...Unless your employer has approved it, do not use it. Check IT department's policy regarding AI. If they do not have a policy, ask they draft one... HIPPA?! Just no, man}{ \rlnk{1n11a3e}.} 
Users are recommended to \emph{Disable telemetry (S3)} to prevent data leakage, and to employ \emph{Tool and Extension monitoring (S4)} and \emph{Logging (S5)} to detect unauthorized behavior and ensure policy compliance.

\observation{Configuration Management (33\%) focuses on securing IDE configurations, improving settings, disabling telemetry, and enforcing compliance checks.}


\noindent\ul{\textbf{Code Governance.}} \emph{Code Governance} requires \emph{Manual Code Verification (S6)}, discourages full LIDE autonomy, favoring manual review of code changes and explicit approval gates. As a user warned: 
\reddittext{all LLM providers have a warning, YOU should double check the output}{~\rlnk{1lrbov5}.}  Closely related to this, \emph{Use Version Control (S7)} serves as a critical recovery mechanism. Practitioners advocated for frequent, incremental commits to enable rollbacks after unintended deletions or silent corruptions: \reddittext{You shouldn't have had 30,000 lines of code in something not backed up...Sync to GitHub.}{~\rlnk{1olmaj3}.}

\observation{Code Governance (31\%) can be achieved by manual code verification and disciplined use of version control to ensure accountability, auditability, and recovery from AI-induced errors.}

\noindent\ul{\textbf{Data Protection and Privacy Control.}} This category details defensive strategies against unauthorized access. The primary strategy, \emph{Sensitive File Protection (S8)}, emphasizes secret isolation via \texttt{.env} files and enforced exclusion rules (\texttt{.gitignore}) to block AI access to hardcoded credentials. For example, in reply to a \emph{.env} related issue, one practitioner suggested: \reddittext{You should have different env for local dev and production, don't trust any tool or AI...}{\rlnk{1l3zhdr}.}


Similarly, \emph{Memory/Context Isolation (S9)} guides for restricting cross-session context sharing to prevent memory mechanisms from retaining sensitive data beyond its intended scope. These strategies reflect a \emph{zero-trust} posture, where practitioners prioritize constraints over built-in LIDE protections.


\observation{Protecting sensitive files and memory isolation (13\%) are treated as necessary countermeasures against inadvertent data exposure rather than optional safeguards.}

\noindent\ul{\textbf{Isolation.}} The \emph{Isolation} strategy treats AI agents as actors requiring explicit containment. The primary strategy, \emph{Sandboxing (S10)}, mandates executing AI-generated code within isolated environments like containers, VMs, or devcontainers to prevent system-wide harm. Following a report where an agent attempted to delete a home folder, a practitioner advised: \reddittext{...you should always run these agents in a VM of some sort, not your host machine}{\rlnk{1ordvv6}.}


\noindent Practitioners also advocate for \emph{Use Local LLM (S11)} to maintain data sovereignty and eliminate external API data transmission. This approach addresses risks regarding opaque vendor retention policies and proprietary code exposure, prioritizing local execution to ensure total control over sensitive data flows.

\observation{Sandboxing or running local LLMs helps developers limit risks from AI-generated code, emphasizing containment rather than trust in AI (13\%).}




\noindent\ul{\textbf{External Guidance.}} \emph{External Guidance} emphasizes leveraging official resources to mitigate the opacity of rapidly evolving LIDE behaviors. The primary strategy, \emph{Consult Vendors (S12)}, involves clarifying security guarantees and resolving functional ambiguities. In response to data privacy issues, practitioners recommended: \reddittext{Request their API data privacy policy.}{\rlnk{13b4u1v}.} Complementing vendor engagement, \emph{Refer to Documentation (S13)} focuses on identifying secure configuration options and intended tool behaviors. These strategies underscore the necessity of authoritative guidance and community reporting to reduce uncertainty and ensure informed LIDE adoption in complex development environments.

\observation{Developers favor following vendor-provided guidelines over relying on built-in guarantees for securing LIDEs (9\%).}




\begin{figure}
\centering
\includegraphics[width=\linewidth]{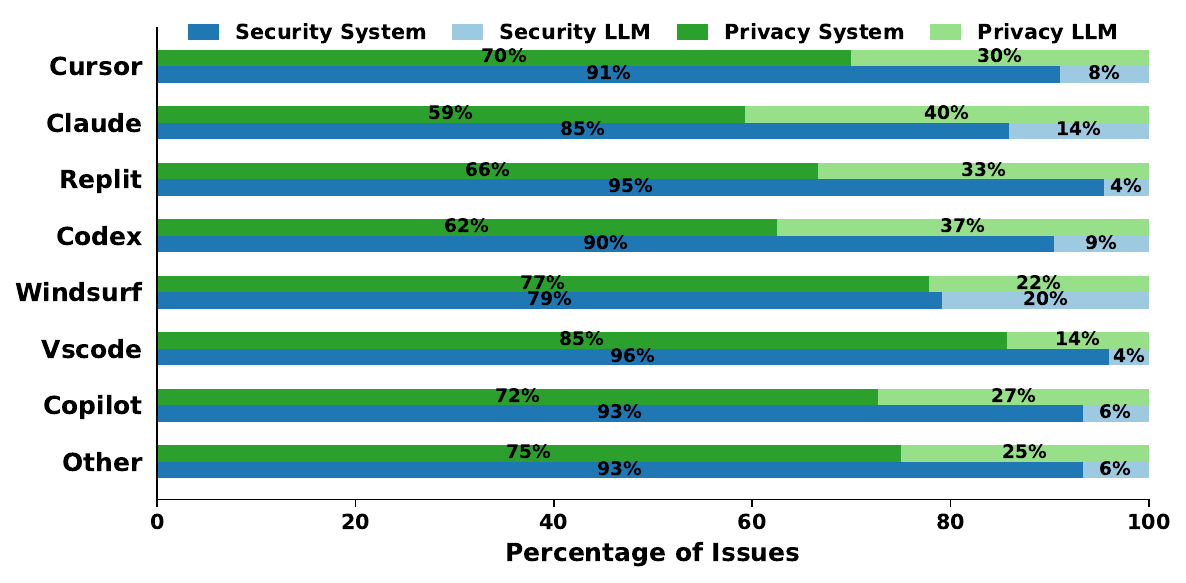}
\caption{\added{LIDEs mentioned in System and LLM Level Issues}}
\label{fig:ide_issues_system_llm}
\end{figure}

\section{Recommendations from Lessons Learned}\label{sec:recommendations}
We offer recommendations to design security- and privacy-aware LIDEs based on the lessons learned from our study.

\noindent\ul{\textbf{System Issues vs LLM Issues.}} While LLM safety has been widely studied, considerably less attention has been paid to the safety implications that emerge from integrating LLMs into IDEs. Many security and privacy failures arise not solely from the LLM’s reasoning behavior, but from how its outputs are interpreted, constrained, and executed by LIDEs. 
\fig\ref{fig:ide_issues_system_llm} shows that most issues discussed in Reddit posts are attributable to insecure integration logic rather than to the mere presence of an LLM. Overall, 7 out of 10 security and privacy issues were system-level issues, and 2 were LLM-level issues (see \href{https://github.com/paper-submission-0/IDESecurityPrivacy/tree/main?tab=readme-ov-file#b-details-on-system-vs-llm-level-issues}{ replication package}); the other one belongs to both categories. For example, unauthorized file access issues were prevalent across all LIDEs. These incidents typically occurred when the LIDE executed LLM-generated actions without enforcing developer-specified access controls. Therefore, LIDE designers need to ensure that proper security and privacy principles are adopted in their system design, irrespective of their usage of an LLM.

\newcommand{\rot}[1]{\rotatebox{90}{\scriptsize#1}} 
\begin{table}
\centering
\scriptsize

\caption{Feature support matrix for the studied LIDEs. A shaded cell indicates IDE support for a feature.}
\label{tab:feature_ide_matrix}
\definecolor{feat}{RGB}{220,230,255}
\resizebox{\linewidth}{!}{
\begin{tabular}{p{.22\textwidth} |cccccccccccccccc}
\toprule
\renewcommand{\arraystretch}{0.7}

\textbf{Feature $\downarrow$ | LIDE $\rightarrow$} &
\rot{\textbf{Aider}} &
\rot{\textbf{Bolt}} &
\rot{\textbf{Codeium}} &
\rot{\textbf{Cline}} &
\rot{\textbf{Cursor}} &
\rot{\textbf{Copilot}} &
\rot{\textbf{JetBrains}} &
\rot{\textbf{Kilo}} &
\rot{\textbf{Kiro}} &
\rot{\textbf{Replit}} &
\rot{\textbf{Roo}} &
\rot{\textbf{Trae}} &
\rot{\textbf{VS Code}} &
\rot{\textbf{Warp}} &
\rot{\textbf{Windsurf}} &
\rot{\textbf{Zed}} \\

\midrule

DEBUG - Code/Bug Debugging &
\cellcolor{feat} & \cellcolor{feat} & \cellcolor{feat} & \cellcolor{feat} & \cellcolor{feat} &
\cellcolor{feat} & \cellcolor{feat} & \cellcolor{feat} & \cellcolor{feat} &
\cellcolor{feat} & \cellcolor{feat} & \cellcolor{feat} & \cellcolor{feat} &
\cellcolor{feat} & \cellcolor{feat} & \cellcolor{feat} \\
\midrule
DOCUMENT - Code Documentation &
\cellcolor{feat} & \cellcolor{feat} & \cellcolor{feat} & \cellcolor{feat} & \cellcolor{feat} &
\cellcolor{feat} & \cellcolor{feat} & \cellcolor{feat} & \cellcolor{feat} &
\cellcolor{feat} & \cellcolor{feat} & \cellcolor{feat} & \cellcolor{feat} &
 & \cellcolor{feat} &  \\
\midrule
REFACT - Code Refactoring &
\cellcolor{feat} & \cellcolor{feat} & \cellcolor{feat} & \cellcolor{feat} & \cellcolor{feat} &
\cellcolor{feat} & \cellcolor{feat} & \cellcolor{feat} & \cellcolor{feat} &
\cellcolor{feat} & \cellcolor{feat} & \cellcolor{feat} & \cellcolor{feat} &
 & \cellcolor{feat} &  \\
\midrule
TEST - Test Generation &
\cellcolor{feat} & \cellcolor{feat} & \cellcolor{feat} & \cellcolor{feat} & \cellcolor{feat} &
\cellcolor{feat} & \cellcolor{feat} & \cellcolor{feat}  & \cellcolor{feat} &
\cellcolor{feat} &  & \cellcolor{feat} & \cellcolor{feat} &
 & \cellcolor{feat} &  \\
\midrule
AUTOCOMPLETE - Code Completion &
\cellcolor{feat} & \cellcolor{feat} & \cellcolor{feat} & \cellcolor{feat} & \cellcolor{feat} &
\cellcolor{feat} & \cellcolor{feat} & \cellcolor{feat} & \cellcolor{feat} &
\cellcolor{feat} & \cellcolor{feat} & \cellcolor{feat} & \cellcolor{feat} &
\cellcolor{feat} & \cellcolor{feat} & \cellcolor{feat} \\
\midrule
SEARCH - Code Search &
\cellcolor{feat} & \cellcolor{feat} & \cellcolor{feat} & \cellcolor{feat} & \cellcolor{feat} &
\cellcolor{feat} & \cellcolor{feat} & \cellcolor{feat} & \cellcolor{feat} &
\cellcolor{feat} & \cellcolor{feat} & \cellcolor{feat} & \cellcolor{feat} &
\cellcolor{feat} & \cellcolor{feat} & \cellcolor{feat} \\
\midrule
EXTOOL - External Tool Execution &
\cellcolor{feat} & \cellcolor{feat} &  & \cellcolor{feat} & \cellcolor{feat} & \cellcolor{feat}
 & \cellcolor{feat} & \cellcolor{feat} &  &
 & \cellcolor{feat} & \cellcolor{feat} & \cellcolor{feat} & \cellcolor{feat}
 & \cellcolor{feat} &  \\
\midrule
MMI - Multi-Modal Input &
 & \cellcolor{feat} &  & \cellcolor{feat} &  \cellcolor{feat} &
\cellcolor{feat} &  &  &  &
 &  &  & \cellcolor{feat} &
 &  &  \\

\bottomrule
\end{tabular}
}
\end{table}
\begin{figure}
    \centering
    \includegraphics[width=\linewidth]{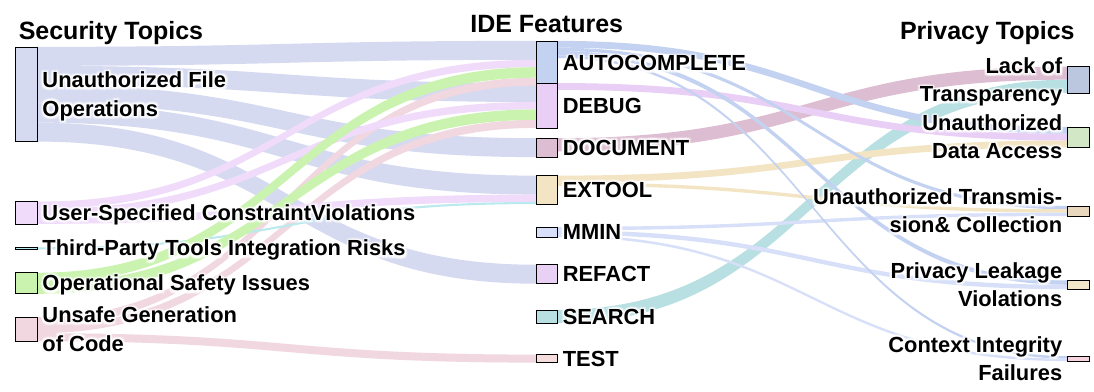}
    \caption{\added{Mapping of AI-assisted IDE features to observed security and privacy issues.}}
    \label{fig:security_privacy_mapping_feature}
\end{figure}


\noindent\ul{\textbf{Balancing Features vs S\&P Guardrails.}} Table~\ref{tab:feature_ide_matrix} outlines the features supported by LIDEs with AI/LLM-enabled capabilities (see \href{https://github.com/paper-submission-0/IDESecurityPrivacy/tree/main?tab=readme-ov-file#d-lide-features}{replication package}). Since identifying these features from potentially incomplete documentation poses a threat, we validated official descriptions through hands-on installation and use of popular LIDEs. While these features enhance the development workflow, \emph{AI Debugging (DEBUG)}, \emph{Code Completion (AUTOCOMPLETE)}, and \emph{External Tool Execution (EXTOOL)} are the most prominent contributors to security and privacy issues (see Figure~\ref{fig:security_privacy_mapping_feature}), collectively connected to all identified issues. Since \emph{DEBUG} and \emph{AUTOCOMPLETE} are ubiquitous across major LIDEs, there is a clear empirical link between widely deployed features and observed issues. Although essential for making LIDEs more proactive, these capabilities simultaneously expand the attack surface and deepen the impact of design flaws. LIDE architectures must therefore enforce proactive security and privacy guardrails as a fundamental requirement for LLM-based feature development, ensuring productivity gains do not come at the cost of system integrity.

\noindent\ul{\textbf{Verification-First Code Recommendations.}} The proliferation of LIDEs has inverted the software development lifecycle, shifting the primary developer role from code synthesis to verification. This transition introduces a silent security debt, where unverified AI output bypasses traditional audits and violates user-specified constraints. Our analysis attributes 17.8\% of security vulnerabilities to unsafe generation (O3), including vulnerable logic and hallucinated dependencies, compounded by a 16.5\% rate of explicit constraint violations (O4). While practitioners advocate for treating AI output as untrusted input (S6, S2), the seamless \textit{Auto-apply} features in modern LIDEs often circumvent manual scrutiny. Consequently, we contend that LIDE architectures have to incorporate a native verification layer to validate generated code against security and privacy standards before integration, ensuring both system integrity and developer liability.


\noindent\ul{\textbf{Promises and Perils of Coding Agents.}} Autonomous coding agents significantly enhance developer productivity by automating complex, multi-step engineering workflows. However, this autonomy introduces security risks through expanded system access. Our findings in Figure \ref{fig:security_privacy_mapping_feature} show that External Tool Execution (EXTOOL) is responsible for 5 out of 10 identified security and privacy issues. These vulnerabilities often stem from third-party integration risks (O5). External components leverage broad permissions and opaque data flows to execute unsafe tools with minimal oversight. To mitigate these threats, the community recommends rigorous monitoring of tools and extensions to detect unauthorized behavior (S4). We argue that LIDE designers must establish a formal protocol to ensure agent trustworthiness. Similar to the Android and Apple app stores, LIDEs require a centralized compliance and verification process. This system must vet coding agents and extensions before granting them access to the development environment.




\noindent\ul{\textbf{Balancing Privacy vs. Contextual Integrity.}} Contextual integrity is vital for AI-assisted development, but it often conflicts with privacy preservation. LLM effectiveness relies on a rich context, which frequently contains sensitive data. Transparency issues and ambiguous retention policies account for 45.9\% of all privacy \added{concerns (O6), whereas AI assistants have been observed retrieving secrets or crossing project boundaries without consent (O7--O10)}. These data flows sometimes occur in plaintext, which reduces transparency and user control. LIDE designers must balance these competing needs during context enrichment. To protect user privacy, designers should implement sensitive file protection to prevent AI tools from accessing secrets or credentials (S8). Additionally, context isolation must restrict the sharing of memory across separate sessions or projects (S9). While data sanitization can strip sensitive information from prompts, excessive filtering often degrades model performance. Designers must therefore implement granular filters that preserve logic while masking sensitive literals to ensure secure yet accurate recommendations.



\begin{figure}
    \centering    
    \includegraphics[width=\linewidth]{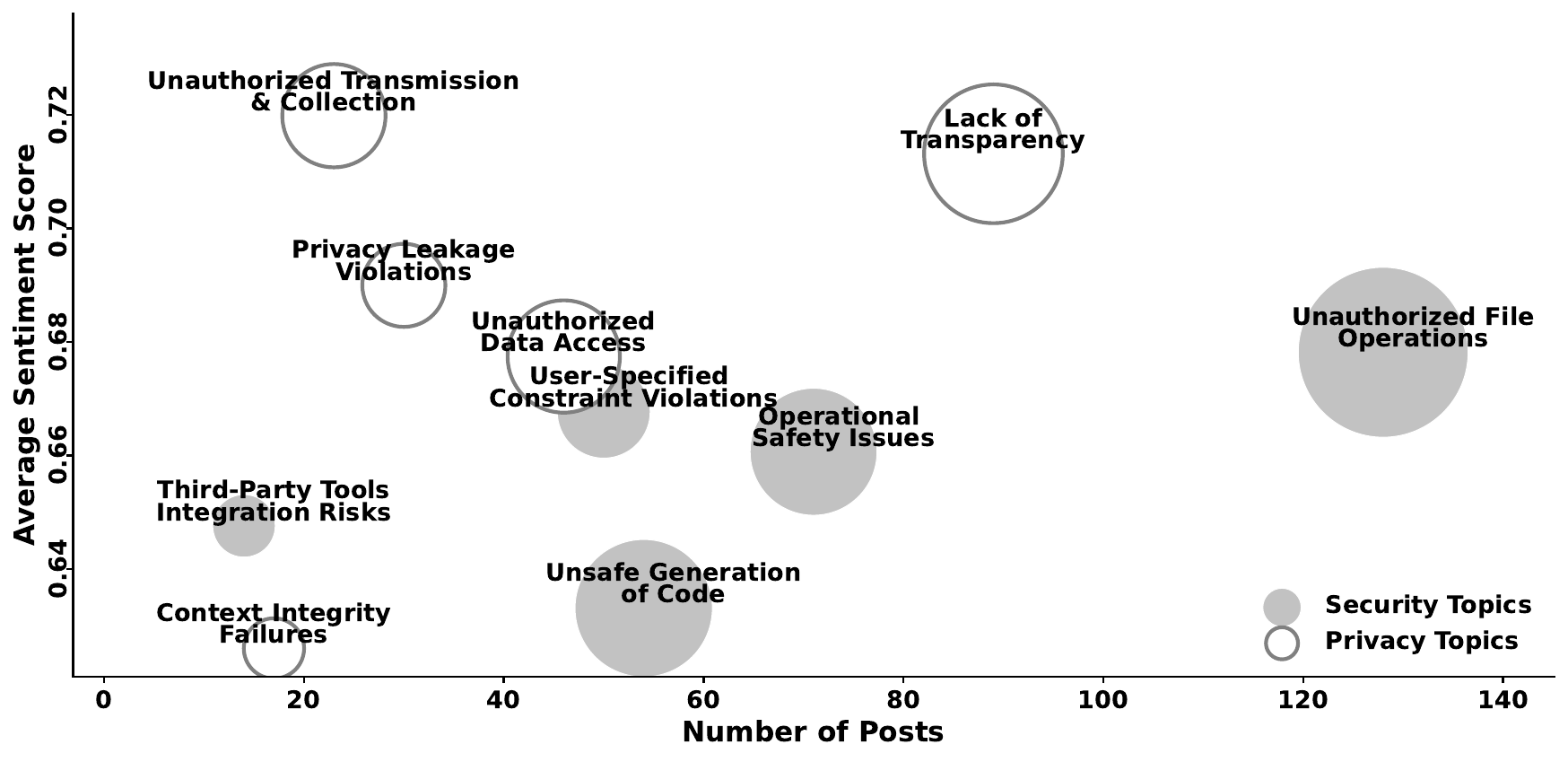}
    \caption{\added{Distribution of security and privacy issues based on user engagement and negative sentiment.}}  

    \label{fig:bubble_chart_combined}
\end{figure}


\noindent\ul{\textbf{Issue Mitigation Prioritization with Developer-in-the-Loop.}} Our observations in \sec\ref{sec:mitigation-strategies} show developers resorting to both proactive and reactive strategies to mitigate reported issues, with comments reflecting prevalent mistrust toward LIDEs. Designers can prioritize incorporating such strategies into their toolkits to ensure more productive, trustworthy ``Dev-LIDE" collaboration. To support this, Figure~\ref{fig:bubble_chart_combined} plots each reported issue as a bubble: higher position (y-axis) reflects more negative developer sentiment (scored via TimeLMs \cite{loureiro2022timelms}), rightward position reflects more posts reporting the issue, and bubble size reflects comment volume. Bubbles higher and further right are thus more critical to developers (e.g., unauthorized file operations, transparency issues). LIDE designers must implement strict security defaults while letting users tune the privacy-performance balance (S1); since developer awareness is essential, the IDE should highlight critical security and privacy trade-offs during installation and configuration.

\begin{figure}
    \centering
    \includegraphics[width=\linewidth]{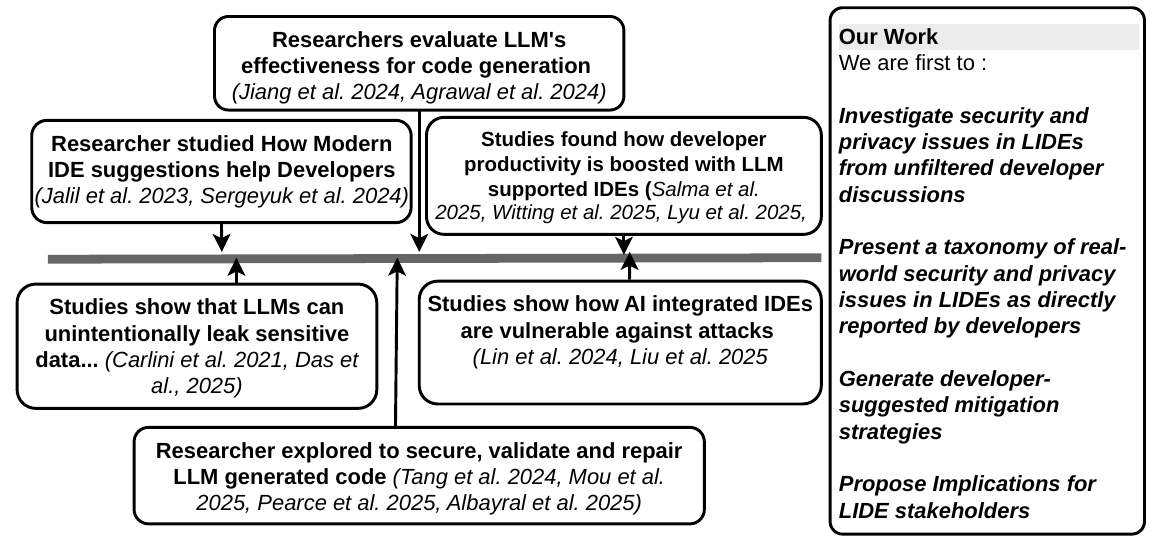}
    \caption{\added{Comparison of our work with related works on LIDEs and the security of LLM-generated code}}
    \label{fig:relatedwork}
\end{figure}
\section{Related Works}

Our related work spans three areas: LLM support in modern IDEs, the impact of LLM-powered IDEs on developer productivity, and the security and privacy implications of LLM-generated code. Figure~\ref{fig:relatedwork} summarizes the related work and highlights our contributions. Unlike the related work, we use unfiltered Reddit discussions to understand security and privacy issues in LIDEs in their real-world use. We summarize selected related works below.

\noindent\ul{\textbf{LLM-native IDEs (LIDEs):}} The emergence of LIDEs marks a shift from rule-based assistance to probabilistic, autonomous execution environments~\cite{Ishankhonov2024TheRO, Sergeyuk2024InIDEHE}. Tools such as GitHub Copilot and Cursor utilize IDE-derived static context~\cite{Li2024EnhancingLC} to enable multi-file reasoning and natural-language-driven task orchestration~\cite{slama2025enhancing}. However, this architecture fundamentally expands the IDE threat model, introducing novel vulnerability classes: \emph{IDEsaster}~\cite{marzouk2025}. By aggregating proprietary source code, secrets, and configurations for external transmission, LIDEs become high-value exfiltration vectors. Recent research exposes stealthy, persistent threats targeting these environments (\emph{Cuckoo Attack})~\cite{liu2025cuckoo} and exploits of overly permissive AI extensions (\emph{UntrustIDE})~\cite{Lin2024UntrustIDEEW}. Complicated by the systemic technical debt inherent in ML-dependent systems~\cite{MLSysTechnicalDebt}, the LIDE architecture becomes a source for unauthorized command execution and large-scale data theft~\cite{hackernews2025}.

\noindent\ul{\textbf{LIDEs and Developer Productivity:}} The integration of LLMs into IDEs has transitioned development from simple autocomplete to end-to-end assisted workflows, yielding 30--70\% time savings across the software lifecycle~\cite{Peng2023TheIO, Li2024EnhancingLC, slama2025enhancing, wittig2025impact, jalil2023transformative}. To reliably measure these gains, execution-based frameworks now prioritize in-environment metrics over traditional text-based benchmarks~\cite{Agarwal2024CopilotEH}. IDEs leverage LLM performance by providing repository-level static context and file references, which facilitates refactoring tasks like \emph{Extract Method} and \emph{Move Method} while enforcing static safety filters~\cite{Ishankhonov2024TheRO, Li2024EnhancingLC, Pomian2024NextGenerationRC, Bellur2025TogetherWA}. However, rapid code generation creates a \emph{validation bottleneck} where productivity hinges on human verification~\cite{Tang2024DeveloperBI}. Behavioral and eye-tracking studies reveal that syntactically polished AI output often induces over-reliance, leading to overlooked logical errors and increased cognitive workloads during inspection phases~\cite{Tang2024DeveloperBI}. Consequently, true productivity assessments must balance speed-to-completion with the mental effort required for rigorous verification and code maintainability~\cite{jalil2023transformative, wittig2025impact, Agarwal2024CopilotEH}.

\noindent\ul{\textbf{Security in LLM-Generated Code:}} 
LLM-generated code frequently mirrors insecure patterns from training data, reproducing OWASP Top 10 vulnerabilities when prompted with legacy or insecure contexts~\cite{pearce2022asleep, jiang2024survey, carlini2021extracting, das2025security}. Newer models like CodeLlama and DeepSeek-Coder show greater functional capability, but their security posture remains inconsistent and domain-dependent~\cite{Li2026PerformanceAO, Taeb2024AssessingTE}—particularly acute in specialized sectors, where LLM-generated PHP code often lacks proper input validation~\cite{Toth2024LLMsIW} and banking logic frequently fails to implement required financial security protocols~\cite{Albayrak2025EvaluatingSA}. These findings reveal a persistent gap where syntactically polished output masks logical flaws, increasing systemic risk to the software supply chain~\cite{Abiha2025UnderstandingCQ, pearce2022asleep}. Current assistants thus cannot be trusted for production-grade software, as they prioritize completion speed over secure logic~\cite{Mou2025CanYR}, necessitating a shift where the developer moves from primary author to security auditor—a high-cognitive-effort role required to catch subtle vulnerabilities and ensure productivity gains do not compromise system integrity~\cite{Mou2025CanYR, Toth2024LLMsIW, Albayrak2025EvaluatingSA}.




\section{\added{Threats to Validity}}

\added{
\textbf{Construct validity} may be affected by inferring security/privacy concerns from informal Reddit posts lacking configuration detail or root-cause evidence; results are thus developer-reported, not confirmed vulnerabilities. We therefore interpret our results as developer-reported issues rather than independently confirmed vulnerabilities in every case. We mitigated this via manual validation of all posts, preserved Reddit identifiers, reproduction attempts to selected posts, and a clarified security/privacy boundary from the end-user perspective. \textbf{Internal validity} risks LLM-filtering misclassification, especially false negatives for implicit issues; we mitigated this by freezing the filtering prompt, evaluating it on a balanced unseen validation, manually validating all LLM-positive candidates, and auditing LLM-negative samples. \textbf{Taxonomy validity} may suffer from overlapping/multi-label categories, addressed by defining boundaries via action, asset, and mitigation implication, with multi-labeling only for explicitly distinct issues. \textbf{External and temporal validity} is limited by reliance on Reddit data, which may not represent enterprise use, proprietary deployments, or developers who do not publicly report issues; Counts reflect discussion patterns, not vulnerability rates or trends. \textbf{Reliability} may vary with evolving discussions; we report coder agreement, resolve disagreements by consensus, and share all scripts, prompts, codebooks, and artifacts. \textbf{Data Availability}: To ensure replication, we share all collected Reddit data. This mitigates the data availability threats that arise when/if Reddit removes/edits the posts.
}

\section{Concluding Remarks}
We present \added{32} security and privacy issues reported by developers in Reddit against popular LLM-native IDEs, aka LIDEs. We observe prevalent mistrust among the developers against the LIDEs, as they attempted 13 different strategies to mitigate the occurrence of such risks. We offer six recommendations to support a more secure and privacy-aware design of LIDEs. 
\begin{takeaway}
\textbf{Message to LIDEs Designers}:  LLM outputs should undergo extensive automated verification before being recommended to developers. The LIDEs designers must treat a LIDE as a software system first, instead of developing it as simply a medium of LLM outputs. Easy access to execution logs and task planning is needed to improve transparency and traceability.
\end{takeaway}
\begin{takeaway}
\textbf{Message to LIDEs Users}: Developers should verify LIDE-suggested code changes before approval. They must follow established software and security engineering practices such as code review rather than ``vibecoding'' only. Evidence suggests that vibecoding is prone to insecure programming~\cite{databrick_vibecoding}.
\end{takeaway}
\begin{takeaway}
\textbf{Message to LIDEs Researchers}:  The academic and industrial researchers \added{can} produce benchmarks to enable robust testing of LIDEs, by taking inspiration from LLM coding benchmarks like SWE-bench \cite{jimenez2023swe}. 
\end{takeaway}

\section{Data Availability Statement}
\label{sec:data-availability}
\added{The replication package with data, code, and supplementary discussion (appendix) is available at
\url{https://zenodo.org/records/21381379}}


\bibliographystyle{ACM-Reference-Format}
\bibliography{main}



\end{document}